# Interest Rates and Inflation
Michael Coopersmith
University of Virginia

Introduction

For some time now I have been intrigued by the apparent relationship between the behavior of various interest rates and that of inflation. As a physicist, I view economics the same way I view physical systems. Namely, describe them by a model which is amenable to mathematical analysis capable of producing real world (testable) results. The following model appears at first glance to be grossly oversimplified; so much so that its connection with the real world should be cast in doubt. We shall see, however, that this is clearly not the case and that its predictions regarding the macroscopic behavior of our economic system arise from a remarkably simple explanation.

The Model

Consider an economic system consisting of just two components, henceforth referred to as manufacturers (M) and bankers (B). The manufacturer earns a profit by producing N objects per year at a cost per year of N×c (the cost per object) and selling them for N×p (the price per object) for an income of N×(p-c). The banker likewise earns an income by producing (borrowing) money (the money supply M) at a low interest $i_S$ (the short term interest rate) and selling (lending) it at a higher interest $i_L$ (the long term interest rate) for an income of M×($i_L - i_S$). We now make the apparently absurd (for the real world) proposition that these two incomes are equal. As we shall see shortly, this assumption (which I shall refer to as the Principle of Equality, the Principle of Equivalence having been preempted by Einstein) will be greatly relaxed, so much so that it should be accepted without dissent. Its purpose here is purely pedagogical; it aids in the visualization of the model.

Our fundamental equation for the Principle of Equality is thus

$$N(p - c) = M(i_L - i_S) \qquad (1)$$

where all of the quantities as defined above are functions of time, t. In the interest of brevity we leave out the variable t at present. Now the producer and consumer price indices are respectively $p/p_o$ and $c/c_o$ where the subscript o indicates an arbitrary fixed time $t_o$. The corresponding inflation rates are the logarithmic derivatives of these indices

$$I_p = \frac{\frac{d}{dt}(\frac{p}{p_o})}{\frac{p}{p_o}} = \frac{d\ln p}{dt} \qquad (2)$$

and

$$I_c = \frac{\frac{d}{dt}(\frac{c}{c_o})}{\frac{c}{c_o}} = \frac{d\ln c}{dt} \qquad (3)$$

Unfortunately, we cannot simply take the logarithmic derivative of Eq. (1) to get an equation involving the inflation rates. But if we make the very reasonable assumption that p and c are linearly related (which seems to be true in a free market system) then there is only one inflation rate which can be written

$$I = \frac{d\ln p}{dt} = \frac{d\ln c}{dt} = \frac{d\ln(p-c)}{dt} \qquad (4)$$

Assuming for the moment that M and N are constant we have from Eq.(1) and Eq.(4)

$$I = \frac{d\ln(i_L - i_S)}{dt} \qquad (5)$$

This equation will be used as the basis for the analyses of various economic conditions in the remainder of this paper. It is immediately evident that Eq.(5) holds for a much more relaxed assumption than Eq.(1) for the incomes of our two components (manufacturers and bankers) need merely be proportional and the more general equation for M and N being time dependent will hold as well. This will be elaborated on in the next section.

Real World Considerations

Although Eq.(5) appears to be a reasonable consequence of our model, the model itself does not take into account a very important aspect of the real world which is true as much for processes in the realm of economics as it is for physics. This is the finite speed of such processes. In physics this caused great consternation after Newton's theory of gravitation was presented because it inherently contained the concept of "action at a distance" or instantaneous cause and effect. It wasn't until Maxwell's electromagnetic equations were solved for moving electric and magnetic sources that the concept of retarded potentials was formed with the speed of light inherently leading to a delay in the effect of a cause. In the present case we must modify Eq.(5) to take account of the time it takes for a change in interest rates to affect the cost of goods and services which are represented in the inflation rate. Equation (5) thus becomes

$$I(t+t_o) = \frac{d\ln(i_L(t) - i_S(t))}{dt} \qquad (6)$$

Where we have used the standard mathematical notation for the argument of a function and $t_o$ is now the delay time, generally taken to be six months to a year. Finally, including N and M in Eq.(6) we get

$$I(t+t_o) + \frac{d\ln N(t+t_o)}{dt} = \frac{d\ln(i_L(t) - i_S(t))}{dt} + \frac{d\ln M(t)}{dt} \qquad (7)$$

We will use Eq.(7) in the next section to analyze some past and present macroeconomic situations to (a) demonstrate its validity and (b) make recommendations for future monetary policy.

Note: Eq.(7) could be said to arise from a modified Principle of Equality (Proportionality), namely

$$N(t + t_o)(p(t + t_o) - c(t + t_o)) = M(t)(i_L(t) - i_S(t)) \qquad (8)$$

which states that the income of manufacturers at time t+t$_o$ is equal or more generally proportional to that of bankers at time t.

Consequences of the Model and Comparison with Known Situations

For simplicity let us start with the case where N and M are both constant in time. It is immediately evident that the inflation rate cannot remain at a given level much less increase because this would require the interest rate difference to increase indefinitely. Specifically, suppose the inflation rate remains constant for a period of time. In order for this to happen the rate of change of the logarithm of the difference of the interest rates must be constant which implies that the form of the interest rate difference be given by

$$i_L - i_S = i_o e^{It} \qquad (9)$$

Where $i_o$ is the interest rate difference at some arbitrary time t=0 and I is the assumed (constant) inflation rate. This actually occurred in the early 1970's to early 1980's when the rate of inflation rose rapidly to around 13.5% before decreasing just as rapidly to its normal level of around 3%. The reason is evident from Eq. (9). Starting at t=0 (1974) with an interest rate difference of , say 3% and an inflation rate of I=10%, the interest rate difference would have almost tripled to an intolerable level in ten years (e=2.718). I cannot emphasize strongly enough that even though Eq. (9) shows exponential growth of the interest rate difference, "exponential" being synonymous in popular parlance with "very large", for values of the exponent (in this case It) small compared with unity the growth is actually linear since $e^x \approx 1 + x$. It is only when $x \approx 1$ that the growth becomes very large (the knee of the curve). Thus in the early stages of interest rate difference growth (which is the situation we find ourselves in at present) it is nearly impossible to tell just how large the logarithmic derivative (and hence the inflation rate) will be.

We now consider the case where the money supply, M, is changing. (Changes in N can be ignored because their effect can be included in M). In the long run the contribution to the inflation rate from the difference of interest rates must be close to zero to prevent runaway exponential growth of these interest rates. The roughly constant (approximately 3%) historical inflation rate is therefore a consequence of the ever increasing money supply which (along with population growth) is essentially exponential.

Discussion

The result of the above analysis is reminiscent of the two-rate scheme originally proposed by Thornton and subsequently reformulated and elaborated upon by Wicksell[1]. There is, however, an essential difference. In the Thornton-Wicksell scheme, once having postulated that inflation is related to a difference of two chosen interest rates, a detailed dynamic argument is made for just what this relationship is. In the present case the relationship arises mathematically from a general principle (equality or more generally proportionality of income) with no reference to the dynamics of the process. In fact, the situation is not unlike the familiar cake cutting problem of its division into two equal parts, the solution being one cuts, the other chooses with no mention of the method (dynamics) of the process. The interest rate difference and the inflation rate are simply related by Eq. 7.

The implication of this equation is a rather disturbing restriction on our ability to control the rate of inflation. Although the Fed can determine the short term interest rate, the long term rate and hence the interest rate difference is determined by market forces. If, for example, the short term rate is raised in an attempt to lower inflation by lowering the difference rate, the long term rate may confound the attempt by increasing as well. If, on the other hand, the long term rate should decrease drastically so as to make the difference rate zero or even negative (inverted yield curve) this will lead to negative inflation (stagflation). These and other effects involving specific relations between the long and short term interest rates will be investigated in a subsequent paper using the theory presented here.